\pdfoutput=1
\RequirePackage{ifpdf}
\ifpdf 
\documentclass[pdftex]{sigma}
\else
\documentclass{sigma}
\fi

\numberwithin{equation}{section}

\begin{document}

\allowdisplaybreaks

\renewcommand{\thefootnote}{$\star$}

\renewcommand{\PaperNumber}{001}

\FirstPageHeading

\ArticleName{Numerical Techniques in Loop Quantum Cosmology\footnote{This
paper is a contribution to the Special Issue ``Loop Quantum Gravity and Cosmology''. The full collection is available at \href{http://www.emis.de/journals/SIGMA/LQGC.html}{http://www.emis.de/journals/SIGMA/LQGC.html}}}

\ShortArticleName{Numerical Techniques in Loop Quantum Cosmology}

\AuthorNameForHeading{D.~Brizuela, D.~Cartin and G.~Khanna}

\Author{David BRIZUELA~$^\dag$, Daniel CARTIN~$^\ddag$ and Gaurav KHANNA~$^\S$}

\Address{$^\dag$~Institute for Gravitation and the Cosmos, The Pennsylvania State University,\\
\hphantom{$^\dag$}~104 Davey Lab, University Park, Pennsylvania 16802, USA}
\EmailD{\href{mailto:brizuela@gravity.psu.edu}{brizuela@gravity.psu.edu}}

\Address{$^\ddag$~Naval Academy Preparatory School, 197 Elliot Avenue, Newport, Rhode Island 02841, USA}
\EmailD{\href{mailto:cartin@naps.edu}{cartin@naps.edu}}

\Address{$^\S$~Physics Department, University of Massachusetts at Dartmouth,\\
\hphantom{$^\S$}~North Dartmouth, Massachusetts 02747, USA}
\EmailD{\href{mailto:gkhanna@umassd.edu}{gkhanna@umassd.edu}}

\ArticleDates{Received October 01, 2011, in f\/inal form December 20, 2011; Published online January 02, 2012}

\Abstract{In this article, we review the use of numerical techniques to obtain solutions for the quantum Hamiltonian constraint in loop quantum cosmology (LQC). First, we summarize the basic features of LQC, and describe features of the constraint equations to solve -- generically, these are dif\/ference (rather than dif\/ferential) equations. Important issues such as dif\/fering quantization methods, stability of the solutions, the semi-classical limit, and the relevance of lattice ref\/inement in the dif\/ference equations are discussed. Finally, the cosmological models already considered in the literature are listed, along with typical features in these models and open issues.}

\Classification{83C45; 83Fxx; 83-08}

\renewcommand{\thefootnote}{\arabic{footnote}}
\setcounter{footnote}{0}

\section{Introduction}

One of the main open questions for any feasible theory of quantum gravity
is that of resolution of classical singularities. Due to the lack of
experimental verif\/ication in this context currently, the resolution of classical singularities
as well as the recovery of the appropriate semi-classical behavior are the
key check that any acceptable theory of quantum gravity must overcome.
Therefore, testing these two issues has been the main goal of all the numerical work
performed in the symmetry reduced context of loop quantum cosmology (LQC).

The geometrodynamical quantization of general relativity \cite{kiefer} is assumed
to be the correct semi-classical limit of any quantum theory of gravity.
In this context, the metric (or in the symmetry reduced cosmological context, the scale factor)
is understood as the fundamental variable. Nevertheless, this quantization can not
be fundamental because the quantum states completely follow the classical trajectories
of the universe in the phase space. More precisely, if a state representing the current
classical universe is evolved backwards making use of the Wheeler--DeWitt (WdW) equation,
it ends up in the big bang singularity. As we will explain below, the situation in LQC
is completely dif\/ferent and, in all analyzed models, the resolution of the corresponding
singularity has been reported.

This article is organized as follows: In Section~\ref{section2}, we provide a brief introduction to LQC with
emphasis on the dif\/ferent quantization schemes used. Section~\ref{section3} discusses (numerical analysis inspired)
stability properties of the equations that arise in LQC models. In Section~\ref{section4}, we survey
a number of isotropic and non-isotropic models. We end this article with a brief summary in Section~\ref{section5}.

\section{Essentials of loop quantum cosmology}\label{section2}

\subsection{Phase space variables and quantum operators}
\label{phase}

In this section, we will develop the general setting of LQC models -- in particular, the origin of the Hamiltonian constraint dif\/ference equation.
This will allow us to make some generic comments about solving these constraints using numerical techniques. In LQG, an oriented spatial manifold $M$ is chosen, with a f\/iducial set of spatial triad vectors $e^a _i$ mapping from the internal vector space (with a positive def\/inite metric $q_{ij}$) to the tangent space (with indices $a, b, \dots$) of the manifold $M$\footnote{In most works on LQC, these f\/iducial inputs are designed as, e.g.~$^o e^a _i$, to dif\/ferentiate them from the physical analogues $e^a _i$, which includes measurable functions such as the scale factors~$a_i (\tau)$. However, we do not need this distinction, so we simplify our notation accordingly. For more discussion, see~\cite{AshPawSin06a, AshWil09a}.}. The dual of the triads are the 1-forms $\omega^i _a$, which give a positive def\/inite metric $q_{ab} = q_{ij} \omega^i _a \omega^j _b$ on $M$. For the Bianchi models, these 1-forms satisfy the Maurer--Cartan relation $d \omega^i = (1/2) C^i _{jk} \omega^j \wedge \omega^k$, where $C^i _{jk}$ are the structure constants of the Lie algebra for the particular Bianchi system.

For the following, we consider specif\/ically the case of a type I Bianchi model~\cite{AshWil09a}, but similar results hold for other systems as well. In particular, we use a diagonal form of Bianchi I, so that the physical metric is given in terms of three scale factors $a_i (\tau)$ and the lapse ${\cal N}$ by
\[
	ds^2 = - {\cal N}^2 d \tau^2 + a_1 ^2 (\tau) dx_1 ^2 + a_2 ^2 (\tau) dx_2 ^2 + a_3 ^2 (\tau) dx_3 ^2  .
\]
As with all Bianchi cosmologies, the spatial slices are non-compact and all physical quantities are homogeneous. This raises the need to restrict all integrations to a f\/iducial cell ${\cal V}$, where all edges of the cell align with the coordinate axes $x_i$ and with a chosen metric $q_{ab}$. For Bianchi~I, this metric can be chosen to be f\/lat, and we designate the sides of the cell along the $x_i$ coordinate to be $L_i$; thus, the volume of the cell ${\cal V}$ is given by $V_0 = L_1 L_2 L_3$.
Then, we can write the triads and connection components as
\begin{gather}
\label{E-A-eqns}
	E^a _i = {\tilde p}_i \sqrt{|q|} e^a _i ,\qquad
	A^i _a = {\tilde c}^i \omega^i _a,
\end{gather}
where $q$ is the determinant of the f\/iducial metric $q_{ab}$.
Throughout this paper, there is assumed to be no summation if all repeated indices are covariant or contravariant; the usual Einstein summation convention will hold when there are mixed indices. Thus, there is no summation in either equation listed in~(\ref{E-A-eqns}).
The phase space of the reduced symmetry system is coordinatized by six variables ${\tilde p}_i$, ${\tilde c}^i$, satisfying the Poisson bracket relation
\begin{gather}
\label{symp-V0}
	\{ {\tilde c}^i, {\tilde p}_j \} = 8 \pi G \gamma V_0 ^{-1} \delta^i _j.
\end{gather}
Here $\gamma$ is the Barbero--Immirzi parameter, a quantization ambiguity which has been calculated to be $\gamma \simeq 0.2735$ by matching with the Bekenstein--Hawking black hole entropy formula~\cite{Mei04}.

However, there is a freedom to re-scale\footnote{Although we do not address it explicitly here, there is also the issue of the behavior of the system under parity transformations~\cite{AshPawSin06b, AshWil09b}. The appropriate choice (when there are no fermions) allows an ef\/fective reduction in the parameter space ${\vec m}$ over which the wave function is def\/ined, e.g.\ only $m_i > 0$ for all three parameters $m_i$.} the coordinates independently in Bianchi I as \mbox{$x_i \to \alpha_i x_i$}, under which $\omega^i _a \to \alpha^i \omega^i _a$ and $e^a _i \to (\alpha^i)^{-1} e^a _i$ (no summation on~$i$. Under this re-scaling, the variables ${\tilde p}_i$, ${\tilde c}^j$ will similarly change~-- an undesirable feature, since the quantization of the model should not depend on the f\/iducial structure chosen. Thus, we normalize the va\-riab\-les~${\tilde p}_i$,~${\tilde c}^j$ such that
\begin{gather*}
	p_1 = L_2 L_3 {\tilde p}_1, \qquad
	p_2 = L_1 L_3 {\tilde p}_2, \qquad
	p_3 = L_1 L_2 {\tilde p}_3,
\end{gather*}
and
\begin{gather*}
	c^1 = L_1 {\tilde c}^1, \qquad
	c^2 = L_2 {\tilde c}^2, \qquad
	c^3 = L_3 {\tilde c}^3.
\end{gather*}
Another way to say this is that the presence of the f\/iducial volume $V_0$ gives a one-parameter ambiguity in the symplectic structure (\ref{symp-V0}), which is now f\/ixed by the choice of $c^i$, $p_j$. Therefore, the Poisson brackets become
\begin{gather}
\label{symp-rel}
	\{ c^i, p_j \} = 8 \pi G \gamma \delta^i _j.
\end{gather}
If the structure constants $C^i _{jk}$ are non-zero~-- as in the other Bianchi cosmologies~-- there is the need to normalize them accordingly~\cite{AshWil09b}.

In the full theory, the variables are the f\/luxes of the triads $E^a _i$ and the holonomies $h_e$, def\/ined by a connection $A^i _a$ along an edge $e$, where the latter is given by $h_e (A) = {\cal P} \exp (\int_e A^i _a \tau_i {\dot e}^a dt)$. Here, $\tau_i$ are the generators of the SU(2) Lie algebra, such that $\tau_i \tau_j = (1/2) \epsilon^k _{ij} \tau_k - (1/4) \delta_{ij} \mathbb{I}$ (these are related to the Pauli matrices $\sigma_k$ as $\tau_k=-i\sigma_k/2$), and $\mathbb{I}$ is the unit $2 \times 2$ matrix. The fact there are no quantum operators corresponding to the connection components $c^i$ in LQG is an indicator of the dif\/feomorphism invariance of this theory. To match LQG as much as possible, we exponentiate the connection components appropriately to get the holonomy $h^{(\delta)} _k$ along the coordinate axis $x_k$ with an edge of length $\delta L_k$, given by
\begin{gather*}
	h^{({\delta})} _k=e^{\delta c_k \tau_k} = \cos \left( \frac{\delta c_k}{2} \right) \mathbb{I} + 2 \sin \left( \frac{\delta c_k}{2} \right) \tau_k.
\end{gather*}
Here $\delta$ is a constant value and the holonomy is given in terms of the almost periodic functions $\exp(i \delta c_k)$~-- ``almost'' because $\delta$ can be any real number, not just an integer.

We consider a Hilbert space basis of eigenstates $|m_1, m_2, m_3\rangle \equiv |{\vec m}\rangle$  of the momentum ope\-ra\-tor~$p_i$. Here, the values $m_i \in \mathbb{R}$ may either be the momentum eigenvalues themselves~-- i.e.\ $p_i ({\vec m}) = m_i$, and therefore have the appropriate physical dimensions -- or dimensionless quantities, so that the dimensions are carried within the function $p_i ({\vec m})$, depending on the choice made in a particular physical model. These eigenstates satisfy
\[
	\langle m_1, m_2, m_3 | m_1', m_2', m_3' \rangle = \delta_{m_1, m_1 '} \delta_{m_2, m_2 '} \delta_{m_3, m_3 '} .
\]
Because there is a Kronecker delta on the right-hand side, rather than a Dirac delta distribution, the wave function is made of a countable number of these orthonormal basis states, in the following form
\[
	\psi = \sum_{m_1, m_2, m_3} \psi_{m_1, m_2, m_3} |m_1, m_2, m_3\rangle,
\]
where the usual f\/initeness condition
\[
	\sum_{m_1, m_2, m_3} | \psi_{m_1, m_2, m_3}|^2 < \infty,
\]
is assumed. This quantization is in-equivalent to the standard Schr\"odinger form, and thus to the usual quantum cosmology using the Wheeler--DeWitt equation as a starting point.
When the triad and holonomy are promoted to quantum operators, they act on the states as
\begin{gather*}
	{\hat p}_i |{\vec m} \rangle = p_i ({\vec m}) | {\vec m} \rangle\,, \qquad
	\widehat{\exp (i \delta c^1)} |{\vec m} \rangle = | m_1 + k \delta, m_2, m_3 \rangle,
\end{gather*}
for a constant $k$ and similarly for the other holonomy operators. The function $p_i ({\vec m})$ and $k$ must combine to satisfy the Poisson bracket relation (\ref{symp-rel}); for Bianchi~I, the triad eigenvalue $p_i ({\vec m}) = m_i$ and $k = -8\pi G \gamma \hbar$. Thus, we see that when the Hamiltonian constraint ${\hat C} = {\hat C}_{\rm grav} + {\hat C}_{\rm matter}$ acts on the wave function, since the operators either give functions of ${\vec m}$ or shift the basis state $|{\vec m}\rangle \to |{\vec m} + k{\vec \delta}\rangle$, the equation ${\hat C} \psi = 0$ gives a dif\/ference equation for the coef\/f\/icients $\psi_{m_1, m_2, m_3}$. Up to now, we have considered $\delta$ to be a constant; however, when we consider quantization methods in the following section, we will see it is physically more viable to use a functional form
$\delta = \delta({\vec m})$.

\subsection{Quantization methods}
\label{quant-meth}

We consider here only the gravitational piece $C_{\rm grav}$ of the Hamiltonian constraint, coming from integrating the full constraint over the f\/iducial cell ${\cal V}$,
\[
	C_{\rm grav} = \int_{\cal V} N {\cal H}_{\rm grav} dV.
\]
The Hamiltonian ${\cal H}_{\rm grav}$ depends on the curvature tensor $F^i _{ab}$, so to quantize this tensor, we write it in terms of our classical phase space variables, and then promote them to quantum operators. In the classical theory, this is done by considering a plaquette $\Box_{ij}$~-- a closed loop whose sides are parallel to the coordinate axes $x^i$ and $x^j$~-- and def\/ining
\[
	F^i _{ab} = 2 \lim_{{\rm Ar}_\Box \to 0} {\rm Tr} \left( \frac{h_{\Box_{ij}} - \mathbb{I}}{{\rm Ar}_\Box} \tau^i \right) \omega^i _a \omega^j _b.
\]
Here, ${\rm Ar}_\Box$ is the area of the plaquette and the holonomy $h_{\Box_{ij}}$ is taken as
\[
	h_{\Box_{ij}} = \bigl[ h^{(\delta_j)} _j \bigr]^{-1} \bigl[ h^{(\delta_i)} _i \bigr]^{-1} h^{(\delta_j)} _j h^{(\delta_i)} _i,
\]
where $\delta_i L_i$ is the length of the $i$th edge of the plaquette, measured in the f\/iducial metric $q_{ab}$. In this scheme, the area of the plaquette is shrunk to zero, so the particular choice of loop is irrelevant.

\begin{figure}[t]\centering
	\includegraphics[width=110mm]{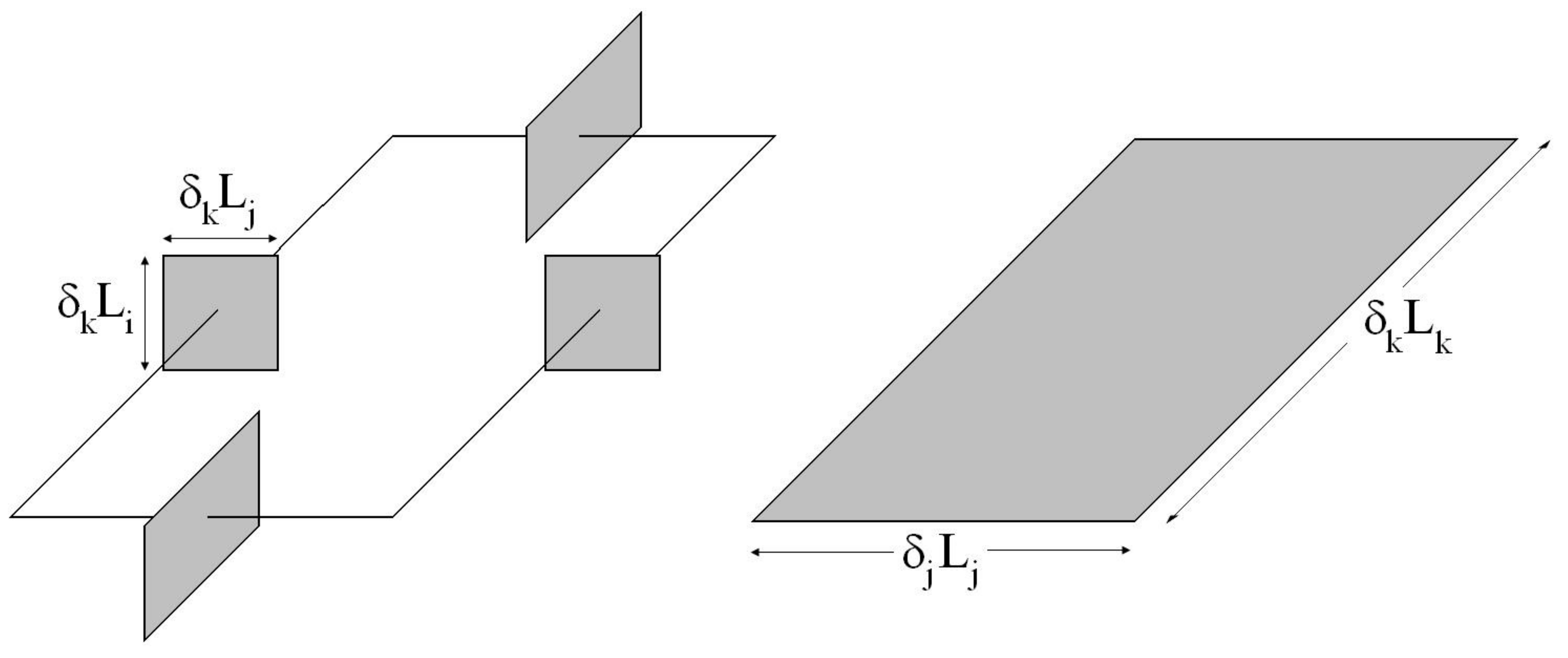}
	\caption{\label{mu-scheme}Illustration of the ${\bar \mu}$ (left picture) and ${\bar \mu}'$ (right) quantization schemes, described in the text. The shaded regions in each diagram indicate the area set equal to the minimum area eigenvalue $\Delta \ell_P ^2$, with the corresponding sides labeled by the edge lengths. Figure adapted from Chiou~\cite{Chi07}.}
\end{figure}

However, this procedure cannot be done in LQG, due to the discreteness of the spectrum for the area operator $\widehat{\rm Ar}$; instead, the plaquette can only be shrunk until its size reaches the minimum area eigenvalue $\Delta \ell_P ^2$ of the area operator, where $\ell_P$ is the Planck length\footnote{From the full theory, we have that the dimensionless number $\Delta = 4 \pi \gamma \sqrt{3}$ although the value $\Delta = 2 \pi \gamma \sqrt{3}$ is sometimes used in earlier works; see the discussion in Appendix B of~\cite{Ash09} for further detail.}. There appears in the literature three separate methods for doing this, with various physical rationales; we discuss these here, and in particular their impact on the eventual dif\/ference equation.
\begin{itemize}\itemsep=0pt

	\item {\it $\mu_0$ scheme:} In this case, the (matrix elements of the) holonomy $h^{(\mu_0)} _j$ is thought of as an eigenstate of the area operator $\widehat{\rm Ar}_j = \widehat{|p_j |}$; the accompanying eigenvalue is set equal to the area gap $\Delta \ell_P ^2$. In other words, with
	\[
		\widehat{\rm Ar}_j h^{(\mu_0)} _j = (C \mu_0) h^{(\mu_0)} _j,
	\]
	for some constant $C$, we let
	\[
		\Delta \ell_P ^2 = C \mu_0 \quad \Rightarrow \quad \mu_0 = \frac{\Delta \ell_P ^2}{C}.
	\]
	This gives a constant value for $\mu_0$; this scheme was used by Ashtekar, Pawlowski and Singh for the $k=0$ isotropic case~\cite{AshPawSin06a}, and Chiou for Bianchi I~\cite{Chi06}. The advantage of this method is that the dif\/ference equation has constant steps, i.e.\ def\/ined on a lattice with even spacing, and therefore is relatively easy to f\/ind solutions. However, this scheme was realized to be problematic in terms of physical predictions~\cite{Ash09}. For instance, as we will explain in Section~\ref{section4}, although the singularity is resolved~-- i.e.\ the cosmological model exhibits a ``bounce'' if a scalar f\/ield is chosen to be the time parameter~-- the critical energy density $\rho_{\rm crit}$ at which this occurs is dependent on the momentum $p_\phi$ of the scalar f\/ield, with $\rho_{\rm crit} \sim p^{-1} _\phi$. Thus, a semi-classical state for the f\/ield, with large $p_\phi$, would have a~correspondingly low value for $\rho_{\rm crit}$; indeed, there is no lower bound for this density, so even densities considered far in the classical regime could be critical densities for the appropriate system.
\end{itemize}
To mitigate these issues, the constants $\delta_i$ associated with each edge of the plaquette can be generalized into functions $\delta_i({\vec m})$ of the parameters. Two possible ways to do this are considered below; we use here the notation of Chiou~\cite{Chi07}. Both of these are dependent on the LQC spin network state~$|{\vec p}\rangle$~-- parameterized by the triad eigenvalues~$p_i ({\vec m})$~-- and its relation to the plaquette chosen.
\begin{itemize}	\itemsep=0pt
	\item {\it ${\bar \mu}$ scheme:} This particular scheme was f\/irst suggested by Chiou~\cite{Chi06}, and it is inspired by the fact that physical areas in LQG are associated with edges between vertices in the spin network. Thus, we imagine the side of the f\/iducial cell ${\cal V}$ normal to the $x_3$ axis to be pierced by $N_3$ edges, each of which is associated with an area and giving a plaquette $\Box_{12}$. Since for a state
$|p_1,p_2,p_3\rangle$ on the kinematical Hilbert space, the area of the face of the f\/iducial cell, orthogonal to the direction $x_i$, is given by $p_i$, we have~\cite{AshWil09a}
	\begin{gather}	
	\label{N3-edges}
		N_3 \Delta \ell_P ^2 = |p_3|.
	\end{gather}
	Since we are assigning each of these edges in the $x_3$ direction to an area parallel to the 1-2 plane, this gives a plaquette with sides of coordinate size $\delta_3 ({\vec p}) L_1$ and $\delta_3 ({\vec p}) L_2$, as in Fig.~\ref{mu-scheme}~\cite{Chi07}. The total area of the 1-2 side must be $L_1 L_2$ in the f\/iducial metric, so that
	\[
		N_3 \delta_3 ^2 ({\vec p}) L_1 L_2 = L_1 L_2.
	\]
	Solving for $\delta_3 ({\vec p})$, and making the same argument for the 1-3 and 2-3 sides, we have that the functions $\delta_i ({\vec p})$ in the Bianchi I model are
	\begin{gather*}
		\delta_1 ({\vec p}) = \ell_P \sqrt{\frac{\Delta}{p_1}}, \qquad
		\delta_2 ({\vec p}) = \ell_P \sqrt{\frac{\Delta}{p_2}}, \qquad
		\delta_3 ({\vec p}) = \ell_P \sqrt{\frac{\Delta}{p_3}}.
	\end{gather*}
	In Bojowald, Cartin and Khanna~\cite{BojCarKha07}, this is the case where the number of vertices is proportional to the transverse area. This method is similar to that of the $\mu_0$ scheme, in that the dif\/ference equation for the Bianchi~I model is def\/ined on a constant spacing lattice, after the appropriate redef\/inition of variables~\cite{Chi06}. However, like the $\mu_0$ scheme, working through the ef\/fective dynamics of the system shows that its physical predictions are dependent on the choice of f\/iducial cell~${\cal V}$~-- in particular, the densities at which a~bounce occurs in one of the three spatial directions~\cite{Chi07}.
		
	\item {\it ${\bar \mu}'$ scheme:} This method was brief\/ly considered by Chiou~\cite{Chi06}, and developed in detail by Ashtekar and Wilson-Ewing~\cite{AshWil09a}. The f\/irst step in this scheme is similar to that of the~${\bar \mu}$ method, namely considering the edges passing through a particular face of the f\/iducial cell ${\cal V}$. Thus, we again have the relation (\ref{N3-edges}) for the relation between the number of edges passing through the 1-2 face and the corresponding value of the triad parameter~$p_3$. However, we now divide this face into plaquettes with lengths $\delta_1 ({\vec p}) L_1$ and $\delta_2 ({\vec p}) L_2$ in the f\/iducial metric as shown in Fig.~\ref{mu-scheme}, with the total area equaling $L_1 L_2$, so that
	\[
		N_3 \delta_1 ({\vec p}) L_1 \delta_2 ({\vec p}) L_2 = L_1 L_2,
	\]
	or
	\[
		\delta_1 ({\vec p}) \delta_2 ({\vec p}) = \frac{\Delta \ell_P ^2}{|p_3|}.
	\]
	If we iterate this for the other two planes, we f\/ind for Bianchi I that
	\begin{gather}
	\label{mubar-delta}
		\delta_1 ({\vec p}) = \ell_P \sqrt{\frac{\Delta p_1}{p_2 p_3}}, \qquad
		\delta_2 ({\vec p}) = \ell_P \sqrt{\frac{\Delta p_2}{p_1 p_3}}, \qquad
		\delta_3 ({\vec p}) = \ell_P \sqrt{\frac{\Delta p_3}{p_1 p_2}}.
	\end{gather}
	This is where the number of vertices in a particular direction is proportional to the extension in that direction~\cite{BojCarKha07}. It can be shown that the physical predictions of this version do not depend on the original f\/iducial cell\footnote{Strictly speaking, this lack of dependence on the volume of the f\/iducial cell ${\cal V}$ is true only for the ef\/fective semi-classical equations, but does not hold for the quantum theory. Indeed, looking at holonomy corrections at all orders, one f\/inds the expectation values $\langle {\hat V} \rangle$ and $\langle {\hat V}^2 \rangle$ depend on the f\/iducial cell due to the dif\/ferent scaling properties of various pieces of these functions (see Section~IV of Chiou and Li~\cite{ChiLi09}). However, this is an expected occurrence, as argued by Chiou~\cite{Chi07}, analogous to the conformal anomaly in quantum f\/ield theory; physical input on the ``correct'' size of the f\/iducial cell may obviate the worry about scaling dependence in the ${\bar \mu}$ scheme.}, so the critical density is invariant under changes in any scalar f\/ield used as a time parameter. However, it also results in a dif\/ference equation for the Hamiltonian constraint where the step sizes are parameter-dependent, so the lattice is not made of constant sized steps. From the analogous situation in computational physics, we denote this as a ``lattice ref\/inement'' model.
\end{itemize}
Note that the ${\bar \mu}$ and ${\bar \mu}'$ schemes are equivalent when there is only a single triad parameter, as in the isotropic cases.
In the literature, the $\mu_0$ method is the ``original'' quantization scheme, while~${\bar \mu}'$ is called the ``improved'' quantization. Nevertheless, other schemes are possible in principle; see for instance~\cite{NelSak10}, where isotropic
embeddings of consistent quantizations of anisotropic models are considered.

\section{Suitability of the dif\/ference equation}\label{section3}

When one f\/inds the numerical solution of a dif\/ferential equation, representing the time evolution of the system, there is a good deal of freedom in choosing the particular discretization scheme for this problem, since small dif\/ferences in this scheme will be of higher order, and thus drop out. Thus, such numerical algorithms are chosen to incorporate a variety of nice features -- not only matching the original equation in the continuum limit, but also stability under numerical computation. The situation is the opposite for LQC, where the dif\/ference equation is fundamental, and cannot be altered. However, there is still some ambiguity in the model, dependent on the quantization method used to arrive at the f\/inal equation; the suitability of these equations depends not only on mathematical arguments~\cite{CorSin08} (such as existence of an inner product in the Hilbert space), but also numerical factors, much as the case with dif\/ferential equations.

In this section, we consider several issues important in the process of numerically solving the Hamiltonian constraint for LQC. First, we must ensure that the dif\/ference equation is stable, meaning that solutions ``close'' in the appropriate sense do not exponentially separate as the system is iterated by increasing one of the model parameters. This leads to the notion of von Neumann stability analysis. Next, the quantum constraint equation must have an appropriate semi-classical limit. One must be careful that this limit is unique, which has not been the case for all models~-- the Schwarzschild interior will serve as our example here. Finally, it is helpful to consider changes in variables which decrease the dif\/f\/iculty in solving the dif\/ference equation. A~particular problem, coming from the improved quantizations detailed earlier, is that the lattice spacings are no longer constant, but depend on the model parameters themselves. In some cases, this can be alleviated by choosing a new parameter as a function of the old, where the lattice spacing becomes constant up to higher order terms.

\subsection{von Neumann stability analysis}

In the latter half of the 20th century, techniques were developed to numerically compute solutions for important dif\/ferential equations by discretizing the equation and evolving in time on a lattice where each point represented a specif\/ic space-time location. It was quickly realized that such solutions could grow without bound if a bad choice of discretization or time step was made. Thus was born the idea of stability analysis, to check the algorithm used and ensure solutions remain of f\/inite size. Since the dif\/ference equation in LQC models is f\/ixed for a particular quantization scheme, the role of stability analysis here is to provide input on the viability of the quantization chosen; the assumption is that an unphysical quantum Hamiltonian constraint would have several problems indicating its lack of consistency -- not only on the basis of its predictions, but also from the characteristics of its numerical solutions. This notion has been born out for several models displaying many undesirable features, such as the isotropic subspace of Bianchi IX~\cite{BojDat04}, the $k=1$ Friedmann--Robertson--Walker (FRW) model~\cite{GreUnr04}, and both
locally rotationally symmetric (LRS) Bianchi I and the Schwarzschild interior~\cite{RosJunKha06}.

We summarize here the discussion of Bojowald and Date~\cite{BojDat04}. In their work, they ensure the notion of a stable dif\/ference equation by checking whether two solutions with ``nearby'' initial data will remain close to each other, rather than diverging. For simplicity, we start with a~dif\/ference equation for a sequence $\psi_m$ of one parameter $m$, or
\begin{gather}
\label{gen-diff-eqn}
	\sum_{i = -k} ^k A_i (m) \psi_{m + i} = 0.
\end{gather}
The analogous analysis can be done for multiple parameters as well. We look at small regions in the parameter space, so that the coef\/f\/icients~$A_i (m)$ of the equation can be approximated by constant values~$A_i$. In general, solutions of dif\/ference equations with constant coef\/f\/icients are found by writing down the characteristic equation
\begin{gather}
\label{char-eqn}
	\sum_{i = -k} ^k A_i \lambda^i = 0,
\end{gather}
and solving for the roots $\lambda_a$~\cite{Ela95}. If there are no repeated roots, the solutions are linear combi\-nations of the form $\psi_m=\beta_1\lambda_1^m+\cdots+\beta_{2k+1}\lambda^m_{2k+1}$, with complex constants $\beta_i$.
For a single repeated root $\lambda_a$ of multiplicity $d_a$, the basis is modif\/ied to $\{ \lambda_1 ^m, \lambda_2 ^m, \dots, \lambda_a ^m, m \lambda_a ^m, \dots, m^{d_a - 1} \lambda_a ^m$, $\dots, \lambda_{2k + 1}^m \}$, and similarly for more than one repeated root. As shown by Bojowald and Date, the physical requirement that the dif\/ference equation has the appropriate classical limit means that $\lambda = 1$ is a double root of the characteristic equation.

{\sloppy The need for a stable dif\/ference equation means that two sequences~$\psi_m$ and~$\psi_m '$, star\-ting at a~pa\-rameter value $M$ with initial data $\{ \psi_M, \psi_{M+1}, \dots, \psi_{M + 2k - 1} \}$ and $\{ \psi_M', \psi_{M+1}', \dots$, $\psi_{M + 2k - 1}' \}$, will remain close together if the initial data is close as well. These sets of values and the dif\/fe\-ren\-ce equation (\ref{gen-diff-eqn}) allow one to solve for $\psi_{M + 2k}$, and all subsequent values for $m > M + 2k$. If we think of each set of initial data as vectors ${\vec S} (M)$ and ${\vec S}' (M)$, respectively, in~$\mathbb{C}^{2k}$ and use the Euclidean norm $|| \cdot ||$, then the initial data are close to each other if $|| {\vec S} - {\vec S}' || \le \delta_M$. Then we require that the sequence $\delta \psi_m \equiv \psi_m - \psi_m'$ remains small in size as well; this places conditions on the roots of the characteristic equation~(\ref{char-eqn}). We see this as follows. For small ranges of the parameter $\delta_M$, $\delta \psi_m$ satisf\/ies the same constant coef\/f\/icient dif\/ference equation~(\ref{gen-diff-eqn}) that the individual sequences $\psi_m$ and $\psi_m'$ do. Thus, we can write $\delta \psi_m$ as a linear combination of the basis elements as
\[
	\delta \psi_m = \sum_{a=1} ^{2k} \sum_{r_a = 0} ^{d_a - 1} C_{a, r_a} m^{r_a} \lambda_a ^m,
\]
where $a$ ranges over the separate characteristic roots, and $C_{a, r_a}$ are constant coef\/f\/icients.
For a~given sequence $\psi_m$ and set of initial data ${\vec S} (M)$, we chose a nearby initial data set ${\vec S}' (M)$ such that only one of the values $C_{a, r_a}$ is non-zero. Then as the sequence $\delta \psi_m$ goes from parameter values $M$ to $M + \Delta M$, we have the norm of the dif\/ference between the two sequences goes as
\[
	\delta_{M + \Delta M} \simeq |\lambda_a|^{\Delta M} \left(1 + \frac{r_a \Delta M}{M} + \cdots\right) \delta_M.
\]
In order to keep $\delta \psi_{M + \Delta M}$ within the same bound on the norm as $\delta \psi_M$, we must have all the roots $|\lambda_a| \le 1$.

}

In Appendix~\ref{closed}, we examine the stability of two versions of the $k=1$ FRW model, that of Green and Unruh~\cite{GreUnr04}, the other by Ashtekar, Pawlowski, Singh and Vandersloot~\cite{AshPawSinVan07}. Although the Green and Unruh model is unsatisfactory due to its unphysical predictions, the stability analysis shows that the problem lies with the quantization method and the resulting Hamiltonian constraint, an issue repaired in the work of Ashtekar et al.

\subsection{Semi-classical limits of the dif\/ference equation}

When using these discrete equations to obtain predictions for semi-classical ef\/fects, one must be careful about how the limit is taken. We assume that the model dif\/ference equation is essentially continuous in the semi-classical limit, in order to match with the existing theory of general relativity. In other words, we should obtain the WdW equation as the limiting case of the dif\/ference equation. For LQC, the (dimensionless) quantum parameter $m$ is related to the continuous variable~$p(m)$ by
\begin{gather*}
p (m) = \gamma \ell m,
\end{gather*}
where $\gamma$ is again the Immirzi parameter, and $\ell$ is proportional to the square of the Planck length. Since $p(m)$ is the eigenvalue of the triad operator, it is related to the scale factor functions appearing in the classical metric. One possible way to take the semi-classical limit is letting \mbox{$\gamma \to 0$}; however, since $\gamma$ is a small but f\/ixed constant, we must assume that $\gamma \to 0$, $m \to \infty$ such that~$p(m)$ remains constant. This is what is called the ``pre-classical limit'' in Bojowald and Date~\cite{BojDat04}. This respects the physical input that~$\gamma$ represents. Then, we can approximate the discrete evolution equation (\ref{gen-diff-eqn}) with a dif\/ferential equation for a smooth wave function~$\psi(p(m))$. This can be obtained by writing the wave function $\psi_m$ and the coef\/f\/icients~$A^m _i (\gamma)$ in a~power series in~$\gamma$. Putting these expansions back into the original discrete relation, we f\/ind a~continuous dif\/ferential equation as the leading order contribution. For LQC, the WdW equation should emerge as the~$O(\gamma^0)$ term, with the higher order $\gamma$ terms corresponding to quantum corrections, and all terms with inverse powers of $\gamma$ should cancel out. The conditions to obtain such a~physically reasonable situation are derived by Bojowald and Date~\cite{BojDat04}.

In some sense, taking the limit as described above is an example of f\/inding the semi-classical equation f\/irst, then solving the resulting WdW equation for the wave function. However, we could solve for the wave function f\/irst without appealing to the physical nature of the para\-me\-ters~-- using generating function techniques~\cite{CarKhaBoj04,CarKha05a, CarKha05b, CarKha06} or other methods -- and then pick out those solutions that have the appropriate behavior. This corresponds to taking the asymptotic limit $m \to \infty$ without reference to the parameter $\gamma$. This is the ``pre-classical limit'' as it has been used by Cartin and Khanna~\cite{CarKha05a, CarKha05b, CarKha06}. The leading order term in a $m$ power series will remain a dif\/ference equation. Generally, solutions of dif\/ference equations can have many undesirable properties, such as oscillation in sign with each succeeding step (i.e.\ $\psi_m \simeq (-\lambda)^m$ for a positive constant $\lambda$ in the large $m$ limit) or unstable solutions that increase without bound. Some control over these solutions can be obtained by choosing the proper initial data. However, it is possible that there is not enough free parameters to avoid unphysical behavior.

As an example, we look at the original quantization of the Schwarzschild interior~\cite{AshBoj06, Mod06}. If we write it in terms of the two triad components $p_b$, $p_c$ (because of spherical symmetry), the Hamiltonian constraint is of the form
\begin{gather*}
	  \gamma \ell_P ^2 \delta \biggl( \sqrt{p_c + 2 \gamma \ell_P ^2 \delta} + \sqrt{p_c} \biggr) \bigl[ \psi(p_b + \gamma \ell_P ^2 \delta, p_c + 2 \gamma \ell_P ^2 \delta) - \psi(p_b - \gamma \ell_P ^2 \delta, p_c + 2 \gamma \ell_P ^2 \delta) \bigr]	\\
 {} + \biggl( \sqrt{p_c + \gamma \ell_P ^2 \delta} - \sqrt{p_c - \gamma \ell_P ^2 \delta} \biggr)\bigl[ (p_b + \gamma \ell_P ^2 \delta) \psi(p_b + 2 \gamma \ell_P ^2 \delta, p_c) - 2(1 + 2 \gamma^2 \delta^2) p_b \psi(p_b, p_c) 	\\
	  \qquad {}+ (p_b - \gamma \ell_P ^2 \delta) \psi(p_b - 2\gamma \ell_P ^2 \delta, p_c) \bigr]	\\
	 {} + \gamma \ell_P ^2 \delta \biggl(\! \sqrt{p_c - 2 \gamma \ell_P ^2 \delta} + \sqrt{p_c} \biggr) \bigl[ \psi(p_b - \gamma \ell_P ^2 \delta, p_c - 2 \gamma \ell_P ^2 \delta) - \psi(p_b + \gamma \ell_P ^2 \delta, p_c + 2 \gamma \ell_P ^2 \delta) \bigr] = 0.
\end{gather*}
For most of the terms and coef\/f\/icients of the dif\/ference equation, taking the $p_b, p_c \to \infty$ limit is equivalent to that of $\gamma \ell_P ^2 \delta \to 0$. However, there is one problematic term in the equation, namely,
\[
	T(p_b, p_c) = -2 \biggl( \sqrt{p_c + \gamma \ell_P ^2 \delta} - \sqrt{p_c - \gamma \ell_P ^2 \delta} \biggr) \big(1 + 2 \gamma^2 \delta^2\big) p_b \psi(p_b, p_c).
\]
Specif\/ically, if we look at the quantum gravity limit $\gamma \delta \to 0$, but allowing $p_b$, $p_c$ to remain the same size by requiring $m_1, m_2 \to \infty$ accordingly, then
\[
	\lim_{\gamma \delta \to 0} T(p_b, p_c) = - \big(2 \gamma \delta \ell_P ^2\big) \frac{p_b}{\sqrt{p_c}}
\]
while in the large momentum limit, then we have
\[
	\lim_{p_b, p_c \to \infty} T(p_b, p_c) = - \big(2 \gamma \delta \ell_P ^2\big) \big(1 + 2 \gamma^2 \delta^2\big) \frac{p_b}{\sqrt{p_c}}.
\]
The dif\/ference in the limits means this model has the undesirable feature that the nature of the semi-classical limit changes, depending on how this limit is taken. Thus, the best case is to f\/ind a Hamiltonian constraint and dif\/ference equation where the semi-classical regime is the same, regardless of which limit is used to reach it.

\subsection{Lattice ref\/inement and model reparametrization}
\label{lattice}

The advent of these improved quantization schemes unfortunately leads to dif\/f\/iculties in sol\-ving the dif\/ference equation, which now involves lattice ref\/inement. As we saw above, the ${\bar \mu}$ and ${\bar \mu}'$ schemes both involve lattice steps that are functions of the model parameters ${\vec m}$. If we have a dif\/ference equation of two parameters, say $m_1$ and $m_2$, then generic terms in the Hamiltonian constraint for the wave function $\psi (m_1, m_2)$ will be of the form $f(m_1, m_2) \psi [m_1 \pm \delta_1 (m_1, m_2), m_2 \pm \delta_2 (m_1, m_2)]$ where the step functions $\delta_i (m_1, m_2)$ are non-linear functions of the parameters. These dif\/ference equations are more dif\/f\/icult to solve, since unlike relations with f\/ixed steps, the equation for a given choice $\{ {\vec m} \}$ depends on values of $\psi({\vec m})$ not calculated from any other equations. An example of this is the dif\/ference equation arising from a lattice-ref\/ined Schwarzschild model~\cite{BojCarKha07}, where
\begin{gather}
\label{Schwarz-delta}
	\delta_1 (m_1, m_2) = \frac{2 \delta_0}{\sqrt{|m_2|}}, \qquad
	\delta_2 (m_1, m_2) = \frac{2 \delta_0 \sqrt{|m_2|}}{m_1}.
\end{gather}
Note that $\delta_0$ is a f\/ixed parameter related to the minimum area eigenvalue (i.e.\ $\delta_0 \sim \sqrt{\Delta}$). In this case, the Hamiltonian constraint is of the form
\begin{gather*}
  C^+ (m_i) \big[\psi(m_1 + 2 \delta_0 / \sqrt{|m_2|},  m_2 +  2 \delta_0 \sqrt{|m_2|}/m_1) - \psi(m_1 - 2 \delta_0 / \sqrt{|m_2|},\\
    \qquad m_2 +  2 \delta_0 \sqrt{|m_2|}/m_1)\big]  	
 + C^0 (m_i) \big[(m_1 + 2 \delta_0 / \sqrt{|m_2|}) \psi(m_1 + 4 \delta_0 / \sqrt{|m_2|},  m_2) \\
 \qquad{} - 2(1 + 2 \gamma^2 \delta_0 ^2 / m_2) m_1 \psi(m_1, m_2)	
 + (m_1 - 2 \delta_0 / \sqrt{|m_2|}) \psi(m_1 - 4 \delta_0 / \sqrt{|m_2|},  m_2)\big] 	\\
\qquad  {} + C^- (m_i) \big[\psi(m_1 - 2 \delta_0 / \sqrt{|m_2|},  m_2 -  2 \delta_0 \sqrt{|m_2|}/m_1) - \psi(m_1 + 2 \delta_0 / \sqrt{|m_2|},\\
  \qquad{} m_2 -  2 \delta_0 \sqrt{|m_2|}/m_1)\big] = 0.
\end{gather*}

To solve these types of lattice ref\/ined equations, we consider a change in the parameters used in the model, where the equation in the new variables now has constant size steps~\cite{BojCarKha07, NelSak08a}. This is possible always for one-parameter models, i.e.\ isotropic cosmologies, but may not be feasible for equations with multiple parameters. This can be seen as follows. If the wave function depends on only a single variable $m$, then with non-constant steps~$\delta (m)$, we have terms in the dif\/ference equation of the form $\psi[m \pm \delta (m)]$. We assume the step function $\delta (m)$ is linear in a parameter~$\delta_0$ which we use as our lattice spacing. To get constant step sizes, we must write the wave function~$\psi(m)$ in terms of a new function~$N(m)$, such that
\begin{gather}
\label{lattice-func}
	N[m \pm \delta (m)] = N(m) \pm \delta_0 + O\big(\delta_0 ^2\big).
\end{gather}
This would imply that
\[
	\psi[N(m \pm \delta (m))] \simeq \psi[N(m) \pm \delta_0],
\]
so the values $N(m)$ form a lattice with a constant spacing $\delta_0$.
Taking the Taylor series expansion of the left-hand side of (\ref{lattice-func}) gives that this condition requires
\begin{gather}
\label{N-step-eqn}
	\frac{dN(m)}{dm} \delta(m) = \delta_0 .
\end{gather}
If the function $N(m)$ is such that higher order terms in (\ref{lattice-func}) correspond to higher order in $\hbar$ as well, then they can be dispensed with using the appropriate factor ordering of the Hamiltonian constraint operator~\cite{BojCarKha07}. Nelson and Sakellariadou showed the issue of factor ordering is intimately related to that of lattice ref\/inement by showing that, for the f\/lat isotropic model, there is a unique choice of quantization scheme (the ``improved'' version) where the factor ordering ambiguities disappear at the semi-classical level~\cite{NelSak08b}.

An example of this reparametrization procedure is seen for the f\/lat isotropic model~\cite{AshPawSin06b}. The authors of that work use a geometric argument -- considering integral curves of operators related to the holonomy operator -- to f\/ind the new function $N(m)$ such that $\psi(N(m))$ has a constant size lattice. Here we show the same result is obtained by the method described above. In particular, in this isotropic model, the need to obtain valid physical predictions (much like for the Bianchi I model) makes it natural to have a step function $\delta (m) = 2 / (3K \sqrt{m})$, where $K$ is a purely numerical constant. Solving the equation (\ref{N-step-eqn}) for $N(m)$ with $\delta_0 = 1$ gives $N(m)$ is proportional to the volume eigenvalue associated with~$m$, that is, $v(m) =  K \,{\rm sgn} (m) |m|^{3/2}$. Therefore, if we redef\/ine the wave function as ${\tilde \psi} (v) = \psi(m)$, then the action of the operator with the improved quantization function ${\bar \mu} (m)$ gives equidistance steps.

This method becomes problematic for multiple parameter models. As we have seen with the functions $\delta_i ({\vec m})$~-- for Bianchi I, given for the ${\bar \mu}'$ scheme in (\ref{mubar-delta}), while in the Schwarzschild interior, shown in~(\ref{Schwarz-delta})~-- the step functions $\delta_i$ can depend on all parameters $m_i$. Suppose for specif\/icity we have two parameters. Then, we seek to f\/ind two independent functions $N(m_1, m_2)$ such that
\[
	N[m_1 \pm k_1 \delta_1(m_1, m_2), m_2 \pm k_2 \delta_2 (m_1, m_2)] = N(m_1, m_2) + k(k_1, k_2) \delta_0 + O\big(\delta_0 ^2\big),
\]
for some constants $k_i$, and $k$ a function of $k_i$. This requires that
\[
	\frac{\partial N}{\partial m_1} \delta_1(m_1, m_2) + \frac{\partial N}{\partial m_2} \delta_2 (m_1, m_2) = k \delta_0.
\]
For this is to be true for all $\pm k_1$, $\pm k_2$, we need specif\/ically for the Schwarzschild case that
\[
	\frac{2}{\sqrt{|m_2|}} \frac{\partial N}{\partial m_1} = C_1, \qquad
	\frac{2 \sqrt{|m_2|}}{m_1} \frac{\partial N}{\partial m_2}  = C_2,
\]
for constants $C_i$; these equations only have one independent solution, namely $N(m_1, m_2) \sim m_1 \sqrt{|m_2|}$, i.e.\ the eigenvalue of the volume operator for the model. Therefore we see that it is possible to have constant step sizes in one of the two parameters, but not in both~\cite{BojCarKha07}. This simplif\/ication, however, is useful in putting the dif\/ference equation into a form more tractable for numerical computation~\cite{JoeKha}\footnote{Solving a partial dif\/ference equation with non-uniform stepping is not straightforward, i.e., a naive iterative approach towards computing the solution simply does not work. The reason is that, due to the non-uniform stepping, computing the future value of the solution requires one to know past of values of the same, that happen to be dif\/ferent from the ones computed in previous iterations! However, one can overcome this issue through the use of a {\em local} interpolation formula for the solution, that enables one to compute precisely those values of the solution that are actually needed, thus making it possible to propagate the evolution forward~\cite{SabKha08}. Now, if one is able to transform the equation into a form in which one of the variables has a constant step-size, then the local interpolant is not required for that variable, and one can proceed with the computation through a simple stepping procedure on that variable. This simplif\/ies the computation considerably~\cite{JoeKha}.}.

Beyond this analytical technique to place the original dif\/ference equation into a more tractable form, there are also various methods of numerically solving the equation along the lines of the Taylor expansions outlined above. One possibility is to interpolate the solution between lattice points already calculated, using a least-squares f\/it~\cite{SabKha08} (more generally, the wave function is expanded in a Taylor series of the parameters around these previously computed values~\cite{NelSak08a}) and then use this interpolant to compute the values needed to step the evolution forward.

\section{Examples of LQC models}\label{section4}

\looseness=-1
In this section we will mainly restrict our attention to those studies that have faced the
purely quantum dif\/ference equation via numerical techniques. This has been performed
for a number of homogeneous models: FRW with spatially f\/lat ${\rm k}=0$, closed ${\rm k}=1$ and open ${\rm k}=-1$ topologies, with and without the presence of a cosmological constant and, less systematically, for anisotropic
Bianchi I models and Schwarzschild interior.
The quantization of Bianchi~II~\cite{AshWil09b} and Bianchi~IX~\cite{Wil10} models have also satisfactorily been carried out in LQC, but the numerical analysis is still missing.
In a last subsection we will mention several studies where the ef\/fective analysis has been done,
since we consider this is a very important research area of LQC. On the one hand,
many of the main features of the LQC evolution has been f\/irst derived via this ef\/fective treatment.
On the other hand, this ef\/fective analysis have been applied to a wide variety of
scenarios, such as inf\/lation as well as to some anisotropic and inhomogeneous model,
which permitted to extract physical information through systematic numerical evolutions
that would be dif\/f\/icult to perform in the full quantum setting. Of course, since to a certain extent,
heuristics is used within these approaches, care is needed when interpreting the validity of
the physical results.

\subsection[Flat Friedmann-Robertson-Walker]{Flat Friedmann--Robertson--Walker}

Because of its simplicity and the fact that it describes to a very good approximation the
evolution of our universe, the FRW solution has been the most studied model in
LQC literature. In fact, the f\/irst positive results regarding the singularity resolution in the context
of LQC were due to Bojowald \cite{Boj01, Boj02} for this model. The singularity is indeed resolved
not simply because the evolution equation is discrete and, hence, the
state jumps over the zero point. In fact, the evolution through the classical singular
point is also well def\/ined. Therefore, any initial data can be in principle deterministically
evolved through the singularity. But, in order to know precisely what is ``on the other
side of the singularity'' and what is the precise behavior of the physical system,
one needs to construct Dirac observables and solve the dif\/ference evolution equation.
Here is where one has to resort to numerical methods. The seminal work in this aspect
was performed by Ashtekar, Pawlowski, and Singh (APS) \cite{AshPawSin06a}. Let us comment more
precisely about their approach. In their work, the spatially f\/lat FRW model was analyzed with a matter source of massless scalar f\/ield $\phi$. This model is particularly interesting
for the study of singularity resolution since classically a (big-bang) singularity is
unavoidable. The scalar f\/ield is a monotonic function of the cosmological
time, so it can itself be interpreted as an internal clock instead of the scale factor $a$.
On the one hand, this is a widely used approach in cosmology to describe the evolution
of the system without making reference to any f\/iducial structure or coordinates.
On the other hand, in the closed (${\rm k}=1$) model, although the universe undergoes a
recollapse, the scalar f\/ield $\phi$, contrary to the scale factor $a$, is still a monotonic
function and can be interpreted as a time variable.

In this model, the phase space is coordinatized by the gravitational degree of freedom
$(c, p)$, with Poisson bracket $\{c,p\}=8\pi G/3$, by the scalar f\/ield $\phi$ and by
its conjugate momentum $p_\phi$ with bracket $\{\phi,p_\phi\}=1$. The only constraint
left is the Hamiltonian that, in terms of these variables, takes the the form,
\begin{gather*}
	C=N(C_{\rm grav}+C_{\rm matter}),\\
 {\rm with} \qquad C_{\rm grav}:=-\frac{6}{\gamma^2}c^2\sqrt{|p|},
\qquad {\rm and}\qquad C_{\rm matter}:=8\pi G\frac{p_\phi^2}{|p|^{3/2}}.
\end{gather*}
For this system,
a complete set of classical Dirac observables is provided by the constant of motion
$p_\phi$ and by $p|_{\phi_0}$. The latter determines the value of
$p$ at the instant of time at which $\phi=\phi_0$. This set is complete because
they univocally def\/ine a trajectory on the phase space. Therefore,
in order to extract physical information of the quantum system, the evolution
of these observables must be analyzed.

The quantization of the geometric degrees of freedom is performed by following
the methods of full LQG and therefore, as has been explained in Section~\ref{phase}, making
use of holonomies $N_\mu$ and triads $p$ as fundamental variables. In this way,
the gravitational kinematical Hilbert space is ${\cal H}_{\rm kin}^{\rm grav} = L^2(\mathbb{R}_B,d\mu_B)$,
being $\mathbb{R}_B$ the Bohr compactif\/ication of the real line with the Bohr measure $d\mu_B$.
This Hilbert space is properly spanned by the eigenfunctions $|\mu\rangle$ of the triad operator: $\hat p|\mu\rangle= 8\pi\gamma \ell_{\rm Pl}^2\mu/6|\mu\rangle$. On the other hand, the matter degree of freedom $\phi$ is quantized via a standard
Schr\"odinger representation on a Hilbert space ${\cal H}_{\rm kin}^{\rm matt}:=L^2(\mathbb{R},d\phi)$. Here, the operator
associated to the conjugate momentum of the matter f\/ield $\hat p_\phi$ acts as a
derivative operator $\hat p_\phi=-i\hbar \partial_\phi$.

In this f\/irst study by Ashtekar, Pawlowski and Singh, the lapse was taken equal to unity and the $\mu_0$ scheme was used.
The physical states $\Psi(\mu, \phi)$ were then def\/ined as those that are annihilated by the
Hamiltonian constraint. On the full kinematical Hilbert space, given by the product between
the geometric and matter spaces $L^2(\mathbb{R}_B,B(\mu)d\mu_B)\otimes L^2(\mathbb{R}, d\phi)$,
the constraint can be written as an evolution equation in terms of the internal emergent time $\phi$,
\begin{gather}\label{evolutioneq}
\partial_\phi^2\Psi= -\Theta_0 \Psi ,
\end{gather}
where the dif\/ference operator is given by,
\begin{gather}
\Theta_0\Psi(\mu,\phi):=[B(\mu)]^{-1} \big\{f(\mu)	\Psi(\mu + 4\mu_0,\phi) -[f(\mu)	+f(\mu-4\mu_0)]\Psi(\mu,\phi)\nonumber\\
\hphantom{\Theta_0\Psi(\mu,\phi):=[B(\mu)]^{-1} \big\{}{}
+ f(\mu-4\mu_0)	\Psi(\mu-4\mu_0, \phi)\big\},\label{Theta0}
\end{gather}
for certain functions $f(\mu)$ and $B(\mu)$. In particular,
$B(\mu)$ is proportional to the eigenvalues of the operator $\widehat{1/|p|^{3/2}}$,
which is the only geometric factor that appears on the matter part of the Hamiltonian.
Due to the dif\/ference operator, superselection sectors
appear. Let us  denote by~${\cal L}_\varepsilon$ the lattice of points $\{\varepsilon+4\mu_0\}$ on the~$\mu$ axis.
Therefore, the subspaces ${\cal H}_\varepsilon\in{\cal H}_{\rm kin}$, def\/ined as those states with support only on
${\cal L}_\varepsilon$, are left invariant under the action of the operator~$\Theta_0$ and the Dirac observables.

The positive-frequency solutions to this equation form the physical Hilbert space and they
can be written in an integral form as,
\begin{gather}
\label{solution}
\Psi(\mu,\phi)=\int_{-\infty}^\infty dk \tilde \Psi(k) e_k^{(s)}e^{i\omega(k)\phi}.
\end{gather}
In this expression, $\tilde \Psi(k)$ is an arbitrary function, $\omega(k)^2=\pi G(16k^2+1)/3$,
whereas $e_k^{(s)}$ is a~symmetric linear combination of eigenfunctions of the operator $\Theta_0$
with eigenvalue $k$. These combinations have support on the lattice ${\cal L}_\varepsilon$
and are, by construction, invariant under the triad orientation reversal operation. The physical inner
product for a given value of $\phi=\phi_0$ is,
\begin{gather*}
\langle\Psi_1|\Psi_2\rangle_\varepsilon=\sum_{\mu\in {\cal L}_\varepsilon}B(\mu)\bar\Psi_1(\mu,\phi_0)\Psi_2(\mu,\phi_0),
\end{gather*}
and the action of the Dirac observables is given as,
\begin{gather*}
\widehat{|\mu |_{\phi_0}}\Psi(\mu,\phi)=e^{i{\sqrt{\Theta_0}}(\phi-\phi_0)}| \mu |\Psi(\mu,\phi_0),\qquad
{\rm and}\qquad\hat p_\phi\Psi(\mu,\phi)=-i\hbar\frac{\partial \Psi(\mu,\phi)}{\partial\phi}.
\end{gather*}

In this way, there are two possible methods in order to solve the constraint
equation. On the one hand, the constraint can be understood as an evolution equation in $\phi$.
Hence, choosing initial data
$\Psi(\mu,\phi_0)$ at a given value of the scalar f\/ield $\phi=\phi_0$, the physical state is obtained
by evolving them via equation~(\ref{evolutioneq}). On the other hand, one could obtain the eigenfunctions of
the opera\-tor~$\Theta_0$, construct the appropriate symmetric linear combinations $e^{(s)}_k$,
pick up a~function~$\tilde \Psi(k)$, and perform the integral on the right-hand side of equation~(\ref{solution}).
The advantage of the f\/irst method is that it is not necessary to deal with the eigenfunctions of
$\Theta_0$ but, from a numerical point of view, it is more dif\/f\/icult since a large number of
dif\/ferential equations must be solved. The choice of the free data in each procedure (either as
initial data or as the free function~$\tilde\Psi$) will introduce the physical input on the system.
Obviously, the most interesting set of initial data is given by those that describe a large classical
universe at late times, since this is the current state we observe in our universe. In~\cite{AshPawSin06a}
these both procedures were followed to numerically solve the behavior of the system. Let us comment
on those with some more detail.

For the evolution procedure, three dif\/ferent methods were used to pick up initial conditions.
In the f\/irst one a Gaussian (with respect to the Wheeler--DeWitt inner product) peaked on a~classical
universe was chosen. In the other two, an analytical solution of the Wheeler--DeWitt equation,
that is semi-classical at late times, was used. Since equation~(\ref{evolutioneq}) represents an inf\/inite number
of coupled dif\/ferential equation, one needs to restrict to a f\/inite domain in order to perform
the numerical evolution. The boundary was chosen far away from the peak of the initial wave
packet, and purely outgoing boundary data were chosen there. In addition, for the discretization of
the time derivatives of (\ref{evolutioneq}), the fourth-order adaptive Runge--Kutta method was applied. Regarding the direct evaluation of the integral (\ref{solution}), the function $\tilde\Psi(k)$
was chosen as a Gaussian (with the usual measure) peaked on certain value
$k^\star$ (${\ll}-1$) so that it describes a semi-classical state at a time~$\phi_0$.
The eigenfunctions of $\Theta_0$ were numerically obtained and the integral
(\ref{solution}) was evaluated using fast Fourier transform.

Even if, obviously, the considered dif\/ferent initial data sets led to dif\/ferent quantitative results,
the generic behavior and the main qualitative results were robust independently of the method
of choosing the initial data or the value of the free parameters within those methods. Evolving
backwards, the expectation value of the observables follow the trajectory of their classical
counterparts until a certain minimum value. At this point they depart completely from the
classical behavior and they bounce, keeping always a f\/inite value, and connecting an expanding
with a collapsing branch of the universe. In addition, the state is sharply peaked during the whole
evolution. In this way, the classical singularity is resolved and the
LQC evolution equation connects two classical branches of the universe through a quantum phase.
The bounce happens because the system reaches a point where quantum gravity ef\/fects are
appreciable and make the gravity repulsive. In principle, one would assume that this must
happen for high values of the matter density, since we do not observe any quantum gravity
in our daily experience. Indeed, this fact is one of the drawbacks of the analysis presented
in~\cite{AshPawSin06a} since the matter density at the bounce happens to be inversely proportional
to the expectation value $\langle \hat p_\phi \rangle$, as discussed in Section~\ref{quant-meth}.
Therefore, one could pick up a value of $\langle \hat p_\phi \rangle$ so that the bounce would
occur at very low values of the matter density (and, hence, curvatures). This drawback was
f\/ixed in~\cite{AshPawSin06b} by introducing the so-called improved dynamics.

In this analysis, the $\bar\mu$ quantization scheme was used. Note that in this
homogeneous scenario the $\bar\mu$ and $\bar\mu'$ schemes presented in Section \ref{quant-meth}
coincide. The mathematical structure of the formalism is essentially the same as for the
$\mu_0$ case explained above, but physically the main f\/law of that approach was cured.
Within the improved-dynamics scheme, for all the considered cases, it was obtained that
in a backward evolution the expectation values of the Dirac observables remain peaked
on their classical trajectories (corresponding to an expanding branch of the universe) until
the matter density reaches a critical value of of $\rho_{\rm crit}=0.82 \rho_{Pl}$, where
the system undergoes a bounce. After that, the energy density starts decreasing and, once its ratio
with~$\rho_{\rm crit}$ is small, the system evolves again peaked on a (contracting) classical trajectory.
As already mentioned, the value~$\rho_{\rm crit}$ is the same for all analyzed simulations.
In fact, from the ef\/fective analysis of this system, it is possible to write down a modif\/ied Friedmann
equation which provides an analytical expression for this object: $\rho_{\rm crit}=3/(16\pi^2\gamma^3G^2\hbar)$.
This value is in very good agreement with the one found in the full quantum simulations.

Regarding the mathematical issues of these second analysis \cite{AshPawSin06b}, in order to label the eigenfunctions of
the triad operator, the real parameter $\mu$ is replaced by the af\/f\/ine one $v:=J {\rm sgn}(\mu)|\mu|^{3/2}$,
with $J=2\sqrt{2}/(3\sqrt{3\sqrt{3}})$. In this
representation, the action of the components of the holonomies, $\widehat{e^{i\bar\mu/2}}$, simply shift in a
constant step the states,
\begin{gather*}
\widehat {e^{i\bar{\mu}/2}}|v\rangle=|v +1\rangle.
\end{gather*}
This basis is best suited to the volume operator (associated to the elementary cell):
\begin{gather*}
\hat V |v\rangle = \left( \frac{8\pi\gamma}{6}\right)^{3/2} \frac{|v|}{J} \ell_{\rm Pl}^3 |v\rangle.
\end{gather*}
The quantized Hamiltonian constraint, written as an evolution equation, takes the same form
as in equation (\ref{evolutioneq}), with the same structure for the operator,
\begin{gather*}
\bar{\Theta_0}\Psi(v,\phi):=[C(v)]^{-1}  \{g(v)	\Psi(v + 4) -[g(v)	+g(v-4)]\Psi(v,\phi)
+ g(v-4)	\Psi(v-4, \phi)\},
\end{gather*}
but with functions $g(v)$ and $C(v)$, dif\/fering from $B(\mu)$ and $f(\mu)$ that correspond to $\Theta_0$ (\ref{Theta0}).
The general (positive-frequency) solution to this equation can be written in a form completely
similar to (\ref{solution}), where now one should consider eigenfunctions of the new dif\/ference
operator~$\bar\Theta_0$. Finally, the numerical techniques to perform both the evaluation of that
integral, as well as the direct resolution of the evolution equation, are completely analogous
to the previous case.

Here a brief comment about the work by Laguna \cite{Lag07} is in order, which has been
a source of some confusion (for further discussion, see Section~VIB of~\cite{AshPawSinVan07}). The paper by
Laguna f\/irst deals with the evolution of wave packets under the LQC Hamiltonian constraint
for this isotropic model, and
then compares it to the WdW equation for that system. Finding a dispersion equation for
plane wave solutions of these models, it is shown that the group velocity goes to zero in the
LQC model~-- and thus the quantum bounce occurs as the contraction of the spatial volume
becomes an expansion -- precisely at the critical maximum energy density $\rho_{\rm crit}$
shown previously. However, Laguna then goes on to consider modif\/ications of the original LQC equation,
which has the form of the shallow water equation, that would eliminate the bounce. In this context,
of interest to those working in the f\/ield of computational physics, he essentially adds extra terms
to the equation, none of which are motivated from the original LQC quantization scheme.
In this sense, the elimination of the bounce as a spurious ef\/fect is not applicable to LQC
directly, since it involves going outside the quantization method used to arrive at the
Hamiltonian constraint operator.

Finally, even if this is the simplest model of LQC, there is still no complete consensus on
the precise details of the quantization. Even inside the framework of the improved dynamics
(choosing the $\bar\mu$ scheme) there is freedom in the construction of the quantum
Hamiltonian since, as is usual in any quantization process, in order to represent the classical
constraint it is possible to choose dif\/ferent factor orderings for the (non-commuting) operators.
In addition, the fact that one could choose arbitrarily the lapse function also introduces another
ambiguity and gives rise to dif\/ferent quantization schemes and, hence, to dif\/ferent discrete operators~$\Theta$. In a~recent paper~\cite{MOT11} four of these prescriptions have been analyzed and compared. The f\/irst
prescription is the one due to APS and has already
been presented above. In this case, the lapse is taken equal to unity and the quantum constraint
is symmetrized with respect to holonomy operators. In~\cite{ACS07}, a simpler version of
this scheme was presented in the sense that it allows to carry out the complete analytical
resolution of the system. This is usually known as solvable Loop Quantum Cosmology (sLQC).
Its key feature is that the lapse function is chosen to be $V/(8\pi G)$, what produces a quantum
Hamiltonian constraint free of inverse of triads (volume) operator. The origin of the third analyzed
prescription lies on the analysis of Bianchi~I models \cite{AshWil09a}. As in the APS case, the lapse is also
taken equal to one but, contrary to the previous prescriptions, the operator $\widehat{{\rm sgn} (v)}$
appears in the quantum constraint. This prescription is usually denoted as MMO, due to the
names of the authors of the article~\cite{MMO09}, where it was f\/irst introduced. Finally, in the same
paper~\cite{MOT11} a new prescription (sMMO) is introduced that combines dif\/ferent aspects of the
sLQC and MMO ones.

The system has been analyzed for all the mentioned prescriptions following the route of the construction
of eigenfunctions of the corresponding discrete operator $\Theta$. From a numerical point
of view, it was found that the MMO and sMMO prescriptions are much more ef\/f\/icient. This is due
to the fact that, for these cases, the spectrum of $\Theta$ is non-degenerate: the state is
required to be symmetric with respect to a change $\mu\rightarrow-\mu$ and, hence, the
mentioned eigenfunctions are determined once their value at a given lattice point is f\/ixed.
In addition, nontrivial dif\/ferences between prescriptions have been found, although
they might be non observable in practice. For instance, three
observables (the volume, the Hubble parameter, and the energy density of the scalar f\/ield)
have been analyzed and, in all the cases, the dif\/ference between their expectation values
for dif\/ferent schemes turn out to be smaller than their corresponding dispersion.

\subsection[Closed Friedmann-Robertson-Walker]{Closed Friedmann--Robertson--Walker}

Quantum gravity ef\/fects are supposed to modify the behavior
of the system only at Planck scales in such a way that the classical singularity is
resolved. In addition, they are expected to be suppressed at low curvatures in order
to recover the classical trajectory. Since classically the universe undergoes a recollapse,
the closed FRW model represents a more restrictive testbed model for LQC methods
than the f\/lat case: this recollapse should not be af\/fected by quantum geometry modif\/ications. One of the f\/irst proposals for the quantization of this model was presented in~\cite{Boj02}
in combination with the f\/lat (${\rm k}=0$) model. The problem, as has been explained above, is that
this quantization was proven to be unstable~\cite{BojDat04} for the closed case. Nevertheless, since
this isotropic model can be understood as a particular case of Bianchi~IX~\cite{BoVa},
Green and Unruh~\cite{GreUnr04} made use of techniques developed for the BIanchi~IX model~\cite{BDV04} in order to construct a quantum hamiltonian constraint within the $\mu_0$ scheme.
In this analysis, the numerical resolution of this constraint was also obtained. Even if an inner
product and observables, that would provide a neat physical interpretation of the system,
were missing, the divergent behavior of the solutions at large scales already indicated that
the classical recollapse could not recovered.

In~\cite{AshPawSinVan07}, the $\bar\mu$ scheme was used to quantize the closed model. In this
article some important developments were made that shed light on the concerns explained
above. Physically, the most important result was that, due to the quantum ef\/fects, the classical
(big-bang and big-crunch) singularities are both resolved and replaced by a quantum bounce.
This led to a cyclic universe picture, where the quantum wave function evolves through inf\/initely
many classical cycles. The value of the matter density at the bounce is in agreement with
the critical value $\rho_{\rm crit}$, introduced above for the f\/lat model, (provided that the volume
of the universe reaches a macroscopic size). More importantly, the state remains sharply peaked
on the classical trajectory (within the corresponding cycle) and, thus, undergoes the
recollapse at a maximum volume that is in complete concordance with the predictions of the
classical theory. Therefore, the infrared limit of the theory turns out to be right. From a technical point of view, the analysis of~\cite{AshPawSinVan07} is formulated in terms of connections. Contrary to previous quantizations which, for technical reasons, constructed holonomies in terms of the extrinsic curvature, in this approach the connection was used. This procedure mimics better the approach followed in full LQG.

The overall structure and analysis is completely analogous to the f\/lat case \cite{AshPawSin06b}. The kinematical Hilbert
spaces are again given by ${\cal H}_{\rm kin}$. The Hamiltonian constraint can be written as,
\begin{gather*}
\partial^2_\phi\Psi(v,\phi) = -\bar\Theta_1 \Psi(v,\phi),
\end{gather*}
with a discrete operator $\bar\Theta_1\Psi(v,\phi):=\bar\Theta_0\Psi(v,\phi) + \Omega(v,\bar\mu,r_0)\Psi(v,\phi)$,
so that the scalar f\/ield $\phi$ serves as emergent time. The function $\Omega$ depends on the radius of
the 3-sphere $r_0$ with respect to the f\/iducial metric. The f\/lat ${\rm k}=0$ case is recovered by letting $r_0$
tend to inf\/inity, which makes the function $\Omega$ vanishing. This discrete operator also produces
superselection sectors. The key dif\/ference with the $\bar\Theta_0$ operator is that the eigenvalues
of $\bar\Theta_1$ are discrete (see \cite{SKL07} for an analytical proof). More precisely, eigenfunctions
exist for all values of $\omega$ but, in order to be normalizable, they must decay exponentially in both
limits $v\rightarrow\pm\infty$. And this only happens for certain discrete values of $\omega$.
Numerically these eigenvalues were found following a bisection method
(taking advantage of the fact that the normalizable eigenfunction is always a critical solution
between functions exponentially diverging to $\pm\infty$ as $v\rightarrow\infty$).

\subsection{Open Friedmann--Robertson--Walker}

The study of the open ${\rm k}=-1$ isotropic case has been carried out in~\cite{van07}
within the $\bar\mu$ scheme. The discrete evolution operator $\Theta_{-1}$ is constructed
and the scalar f\/ield is understood as emergent time. Following the procedure introduced
in \cite{AshPawSin06a, AshPawSin06b}, eigenfunctions of the $\Theta_{-1}$ operator
were obtained and then Fourier transformed, in the analogous equations to (\ref{solution}),
in order to get the physical states. The qualitative behavior of the system is completely
similar to the plane ${\rm k}=0$ and closed ${\rm k}=1$ models described above.
Regarding the energy density of the scalar f\/ield at the bounce, an analytical expression
is obtained by ef\/fective means, which coincides with the results of the full
quantum system. This expression depends on the momentum of the scalar
f\/ield $p_\phi$, but it tends to $\rho_{\rm crit}$ for large values of $p_\phi$.

In~\cite{Szu07} some issues, like the choice of the loop to def\/ine the
holonomies, were clarif\/ied, but this analysis is not yet totally satisfactory. First, the def\/ined discrete operator is not essentially self-adjoint. Also, as in the f\/irst analysis of the closed model, the
extrinsic curvature, instead of the connection, is used to construct the holonomies.
This route is taken in order to avoid the technical dif\/f\/iculty of constructing an
appropriate loop. In principle this last drawback could be cured following the
proposal presented in \cite{AshWil09b} for the Bianchi II model, but a complete analysis
is still missing.

\subsection{Cosmological constant}

The inclusion of the cosmological constant has only been considered for FRW spatially
f\/lat models with a matter content of a massless scalar f\/ield that, as in previous models,
also serves as internal time.
In particular, in Appendix~A of \cite{AshPawSin06b}, the main techniques
for the $\bar\mu$ scheme were already put forward and some numerical evolutions of this system were presented.
In all cases the the classical singularity happened to be replaced by a quantum bounce and,
contrary to previous treatments \cite{BaDa05, NPV04}, the correct semi-classical behavior of
the system was recovered. Moreover, the total energy density (def\/ined as the sum between
the energy density corresponding to the matter and to the cosmological constant: $\rho:=\rho_\phi+\Lambda/(8\pi G)$)
at the bounce was found to be the same as in the $\Lambda=0$ cases and, hence,
independent of the value of the cosmological constant.

Already at a classical level, the behavior of the system is completely dif\/ferent depending
on the sign of the cosmological constant. Classically, the universe with $\Lambda<0$ begins
at a big bang singularity at a ``time'' $\phi=-\infty$, expands until the total energy density reaches a
minimum (in fact it vanishes) where, even if we are in the spatially f\/lat case, undergoes a recollapse and f\/inishes
at a big-crunch at $\phi=\infty$. In~\cite{BePa08} a concrete and detail analysis of the quantization
of this model was carried out and the results were similar to the closed ${\rm k}=1$ FRW with a~vanishing~$\Lambda$.
The spectrum of the corresponding discrete operator turns out to be discrete and,
therefore, its eigenvalues are isolated. Hence, a similar bisection method to the ${\rm k}=1$ case was
applied. The quantum system also resolves both initial and f\/inal singularities, replacing
them by quantum bounces happening at a value $\rho=0.82\rho_{\rm Pl}$, which results in
a cyclic universe model. In addition, the classical recollapse was also recovered in complete
agreement with the classical results.

In the classical $\Lambda>0$ model, after the big-bang, the
scale factor of the universe reaches inf\/inity at a f\/inite value of $\phi=\phi_0$. Therefore,
contrary to the $\Lambda<0$ model, in this case the values of the scalar f\/ield are
constrained to a f\/inite interval. In~\cite{KaPa09} Kaminski and Pawlowski constructed
the discrete evolution operator making use of the three quantization prescription commented
above (APS, sLQC, and MMO) and studied its possible self-adjoint extensions. The properties
of this operator happened to be completely dif\/ferent depending on the value of the cosmological
constant. For $0<\Lambda<\Lambda_{\rm crit}$, where $\Lambda_{\rm crit} = 8\pi G \rho_{\rm crit}$ is the
value for which the energy density of the cosmological constant equals the maximum total
density $\rho_{\rm crit}$, the operator admits many extensions; whereas for $\Lambda\geq\Lambda_{\rm crit}$,
the operator turns out to be essentially self-adjoint. Observations favor a positive but small value of the
cosmological constant $\Lambda \ll \Lambda_{\rm crit}$ and the dif\/ferent self-adjoint extensions could give
rise to non-equivalent unitary evolutions. Even so, a recent numerical analysis~\cite{AsPa11}
shows that for the relevant states (which are initially sharply peaked at a classical point of the
phase space) the evolution is quite insensitive to the chosen extension. Furthermore, the
classical initial singularity is resolved. The self-adjoint
extensions permit to follow unitarily the evolution in $\phi$ after the classical divergence at $\phi_0$
and give rise to a recollapse. Thus, a cyclic universe picture is again obtained.

\looseness=-1
In order to f\/inalize this section, we would like to brief\/ly comment on another approach that has
been proposed in order to include the cosmological constant in LQC through the so-called
unimodular gravity~\cite{ChGe10}. At classical level, unimodular gravity is equivalent to
general relativity with the only dif\/ference that allows for a dynamical cosmological constant.
In fact, the momentum conjugate to the cosmological constant can serve as emergent time
and it seems that the quantum theory one obtains is simpler than in the cases with the scalar
f\/ield clock. Remarkably, the main results of singularity resolution and the critical value of
the energy density coincide for both cases.

\subsection{Bianchi I}

Bianchi I model is the simplest non-isotropic model one could consider. Even though,
most of the analysis for this model, as a test case for anisotropic
cosmologies, has been performed at the level of the ef\/fective dynamics
rather than solutions of the Hamiltonian constraint dif\/ference equation.
In part, this is due to the greater complexity of the constraint, with a lattice
having non-constant steps in the $\bar{\mu}'$ quantization scheme. Complete
quantizations have been performed both for a massless scalar f\/ield matter content
\cite{Chi06, AshWil09a} and for vacuum \cite{MMP08, MMW10}.
In this section, we will just mention these two quantizations and the only numerical analysis~\cite{MMP09}, even if not completely systematic,
that has been carried out to analyze the full quantum evolution of this model
in the $\bar\mu'$ scheme.

\looseness=1
As in the previous cases, the scalar f\/ield serves as an internal clock and can
be treated as an emergent time. Making use of this fact, the quantization of this system was f\/irst
carried out in~\cite{Chi06} for both $\mu_0$ and $\bar\mu$ schemes. Nonetheless, this approach suf\/fered from several drawbacks, like the incorrect infrared limit and dependence of the physical results on the choice
of cell~\cite{Szu08}, that were f\/ixed by considering the $\bar\mu'$ quantization \cite{AshWil09a}.
The particular form of the Hamiltonian constraint given in~\cite{AshWil09a} is non-symmetric
in its three parameters due to a variable transformation chosen by the authors; instead of using all three triad parameters~$p_i$ to serve as variables in the dif\/ference equation, one of the three is transformed (as discussed earlier in Section~\ref{lattice}) to a~volume~$v$. Thus, the model is given in terms of two directional triad parameters and the total volume. In addition,
it was shown that the dif\/ference equation def\/ined by that Hamiltonian is unstable to growing
modes~\cite{NelSak09}, although it is unclear whether these modes are eliminated by the inner product.
An open area for
further work is whether this construction is the best choice, or if other schemes (including the symmetric
one) are more viable in a numerical sense.

On the other hand, in~\cite{MMP08} a complete quantization of the vacuum Bianchi I model
was deve\-lo\-ped within the~$\bar\mu$ scheme. Subsequently,
the analysis was also extended to the~$\bar\mu'$ quantization prescription~\cite{MMW10}.
In this approach, in order to have a concept of evolution, the role of the clock must be fulf\/illed
by some geometric degree of freedom. In the context of the quantization given in~\cite{MMP08},
two dif\/ferent variables (a triad coef\/f\/icients and its conjugate momentum) have been tested
numerically~\cite{MMP09}. This numerical analysis has been restricted
to Gaussian states and, due to complicate numerical problems, long term evolutions
could not be checked. Therefore, further numerical analysis
are needed to conclude, for instance, that all semi-classical states remain so
after the bounce. Even though, the bounce picture was
proven to be robust under the new interpretation of time.

\subsection{Ef\/fective models}

The approach given by the ef\/fective equations is based on the so-called geometric quantum
mechanics. In this context, the quantum Hilbert space is considered to be an inf\/inite dimensional
phase space with the symplectic form given by the imaginary part of the inner product. Taking expectation
values one can relate both spaces: each operator on the Hilbert space is mapped to a function on the
phase space. Then, it can be seen that the commutator operation of the Hilbert space is directly
related to the Poisson brackets of the phase space. Since the equations of motion need to be solved
only for the classical degrees of
freedom (that are interpreted as the expectation values of the observables in the
quantum theory) with given quantum geometric corrections, one deals with dif\/ferential equations
and not with discrete equations. In particular, for the case of
homogeneous systems, we only need to deal with ordinary dif\/ferential equations
since the unique dependence of all objects will be the time variable. This fact makes
easier, specially from a computational point of view, to analyze the dif\/ferent models
ef\/fectively rather than facing the full quantum evolution equations.

As far as we know, there is only one analysis \cite{Tav08} in the literature, corresponding to the quantization of
FRW model, that has systematically derived the ef\/fective Hamiltonian and equations of motion
by computing the expectation value of the quantum Hamiltonian operator
for Gaussian coherent states and then performing an asymptotic expansion.
For the rest of the cases,
the ef\/fective equation program has essentially consisted on replacing the operators
in the quantum Hamiltonian constraint by their expectation values. This provides
a f\/irst order approach inasmuch as one disregards all state dependent information.
But it gives rise to a modif\/ied classical Hamiltonian that contains corrections by LQG
ef\/fects, from which the ef\/fective equations of
motion for dif\/ferent variables can be obtained.
The ef\/fective evolution of these variables is expected to mimic the behavior of
the quantum system. In fact, in all the cases presented in this section, that the quantum discrete
equation has been numerically solved, the pure quantum results have been compared with the ef\/fective
ones, and a very precise agreement has been obtained.

Ef\/fective descriptions of quantum FRW are used to intuitively understand dif\/ferent
aspects of the system and generalize the obtained results through a purely quantum analysis.
For instance, the analytical form of $\rho_{\rm crit}$ has been obtained \cite{AshPawSin06a}. In addition,
the presence of a~short phase of superinf\/lation immediately after the bounce was
f\/irst shown for a scalar matter f\/ield in~\cite{Bojowald02}. This result was afterwards generalized
to other matter models \cite{Singh06}. During this superinf\/lationary phase both the Hubble
parameter and its time derivative are positive; thus the acceleration of the universe is faster
than in a de-Sitter spacetime. Furthermore, the resolution of strong singularities has also
been shown throughout this ef\/fective analysis~\cite{Singh0911}. A particularly important
question is whether LQC may leave any testable footprints in the CMB. This issue has been
analyzed also by ef\/fective means in several articles (see, for instance, \cite{TSM03, Hos05, ShHa09, GrBa09, Mie10, BCT11})
with a wide variety of assumptions, but yet no f\/inal consensus has been achieved.

On the other hand, ef\/fective equations have been applied to the analysis of more complicated models,
for which a complete quantum treatment is still out of reach. Regarding the anisotropic models, in~\cite{BoDa04} it has been shown
that the chaotic behavior of the classical Bianchi IX model near the singularity is reduced by quantum ef\/fects
for a~$\mu_0$ scheme. Ef\/fective equations for this model in the $\bar\mu'$ scheme have already been proposed~\cite{Wil10}, but a numerical analysis is still missing.

The ef\/fective dynamics of the vacuum Bianchi I model was f\/irst analyzed in
\cite{Dat05} in the context of $\mu_0$ scheme. This case is simple enough so that the
analytical solution can be achieved, which clearly shows the resolution of the classical
singularity since the volume is not allowed to vanish. For generic matter content, and in the
context of the $\bar\mu$ scheme, the ef\/fective study was carried out in~\cite{ChiVan07}.
In particular, the vacuum, scalar f\/ield and radiation matter contents were analyzed.
For the particular case of a scalar matter f\/ield, in~\cite{Chi07} a comparison between
the $\bar\mu$ and $\bar\mu'$ schemes was presented. This analysis showed that in
both cases the classical singularities are resolved and replaced by big bounces (one for each
spatial direction) but certain key details dif\/fer. In the $\bar\mu$ scheme these bounces
occur when the corresponding directional densities $\rho_I := p_\phi^2/p_I^3$, which depend
on the f\/iducial cell, reach certain critical
value. Whereas, for the $\bar\mu'$ case, the characterization
of the bounce is given by the critical value of the matter density. Thus, in the last
case, all three bounces take place almost simultaneously. The classical collapsing universe
that is obtained ``on the other side of the bounces'' also depends on the quantization scheme.

There are also several ef\/fective analysis of the interior of the Schwarzschild black hole and
even if, as far as we know, none of them is based on a rigorous quantum theory, appealing
results have been obtained. For instance, in~\cite{BoVa07} the singularity resolution was
already obtained for both $\mu_0$ and $\bar\mu'$ quantization schemes.
Constructing on these results,
in a manner parallel to the above-commented Bianchi I case~\cite{Chi07},	
Chiou performed a systematic comparison between the physical consequences of
$\bar{\mu}$ and $\bar{\mu}'$ schemes for this model~\cite{Chi08}. In both cases
the classical singularity is resolved but the f\/inal picture is quite dif\/ferent.
Taking into account an extended Schwarzschild spacetime, in the $\bar\mu$
scheme quantum bounces replace the classical white and
black hole singularities and work as a bridge between their corresponding
regions. On the other hand, in the $\bar\mu'$ case, the classical singularity
is resolved, which implies that the even horizon is dif\/fused. This produces
a baby black hole with a very reduced mass in the consecutive classical cycle.
A~process that continues until the spacetime enters a highly quantum regime.

Regarding cosmological inhomogeneous models, the vacuum Gowdy spacetime (with a three-torus spatial topology) has been numerically analyzed in~\cite{BMP10}
for the $\bar\mu$ scheme.
This ef\/fective treatment is based on the hybrid quantization of the mentioned model~\cite{MMG}, where
dif\/ferent degrees of freedom are treated on a distinct footing. The homogeneous degrees of freedom
are quantized polymerically, whereas a regular Fock quantization is developed for the inhomogeneities.
The main result of this analysis is that the bounce picture is robust in the presence of inhomogeneities.
In addition, a Monte Carlo analysis showed that the amplitude of the inhomogeneities is statistically
conserved through the bounce for highly inhomogeneous universes, whereas it is amplif\/ied for almost
homogeneous cases.

In order to f\/inalize, we would like to brief\/ly mention that a dif\/ferent ef\/fective method
(the so-called truncation method~\cite{AsSi11}) has been used in quantum cosmology~\cite{BoSk06, BSS09}.
Following the geometric quantization mentioned above,
a state can be completely parameterized by the expectation values of the basic operators, like $\hat q$
and $\hat p$ in quantum mechanics, and an inf\/inite number of moments (expectation values of
$\hat q^a\hat p^b$ for all~$a$ and~$b$). The equations of motion for all these variables, which
constitute an inf\/inite system of coupled dif\/ferential equations, is then completely equivalent to the
quantum dynamics of the corresponding wave function. Nonetheless, in certain (semiclassical)
situations, this inf\/inite system can be truncated by keeping terms up to a given order (def\/ined as the sum $a+b$).
In this way,
it is possible to analyze the back-reaction of the high-order moments on the trajectories of the
expectation values. For instance, this approach has been used in~\cite{BHS07} to obtain and analyze the
ef\/fective equations (for both WdW and loop quantizations) of cosmological homogeneous
models with a massive or interacting scalar f\/ield at second order. In contrast to the free scalar f\/ield case, the back
reaction is very relevant in this model since the classical equations for the
expectation values receive corrections due to the back-reaction of f\/luctuations.
The LRS Bianchi~I model was also analyzed in~\cite{BoTs09}.
In this case, the matter content was chosen as an isotropic scalar f\/ield with a negative energy
density given in terms of the anisotropic scale factors. This exotic matter f\/ield makes the
system treatable in terms of physical coherent states. Therefore, a comparison between
the ef\/fective and the purely quantum system was performed and both approaches showed an excellent agreement.
Finally, in~\cite{BBHKM}, in order to systematize the computation of equations of motion for
the moments at very high order, powerful algebraic computational tools
have been constructed. Furthermore, these tools have then been applied to the study the quantum
back reaction on a homogeneous universe with positive cosmological constant.

\section{Summary}\label{section5}

In this article we reviewed a variety of numerical techniques and their application to several relevant LQC models. We commented on the features of the quantum Hamiltonian constraint in the context these models that make these techniques readily applicable and extremely useful, especially for exploring dif\/ferent quantization schemes and examining the semi-classical limit. More specif\/ically, inspired by its role in numerical analysis, we reviewed the concept and application of von Neumann stability to several LQC models. We also reviewed the signif\/icant role played by lattice ref\/inement in the context of the semi-classical behavior of these models. This article also includes a survey of a wide variety of isotropic and anisotropic LQC models and how these numerical techniques have contributed towards a deeper understanding of their features, and a discussion of open issues that require further work to advance this emerging f\/ield.

\appendix

\section{Stability analysis of the closed FRW model}
\label{closed}

\looseness=-1
As an example of stability analysis, we consider here the case of the $k=1$ FRW model, with two incarnations of the quantization method -- the work of Green and Unruh~\cite{GreUnr04} with the old quantization scheme, and Ashtekar, Pawlowski, Singh and Vandersloot~\cite{AshPawSinVan07} with the improved quantization method. In the original Green and Unruh paper, the authors found that their Hamiltonian constraint generically led to solutions that increased in size without bound for large values of the triad parameter $p$, obviously not in tune with the classical recollapse expected for a~closed isotropic model. Their discussion of these results notes the absence of an inner product in their framework to eliminate such unphysical states, but they posited that the lack of a~recollapse indicated the LQC program has less contact with the full LQG theory than previous results had indicated. However, the results of Ashtekar et al., using a Hamiltonian constraint with an improved quantization scheme with lattice ref\/inement, put to rest those doubts by f\/inding both a~bounce in the quantum regime near the classical singularity, and a~recollapse at the appropriate classical scale. In their comparison of the results in the two papers (Section~VIIB in~\cite{AshPawSinVan07}), they point out two dif\/ferences between the constraints~-- the self-adjoint nature of the improved quantum constraint, and the use of a scalar f\/ield by Ashtekar et al.\ as a~time parameter for the evolution.

However, we show here that the Green and Unruh constraint suf\/fers from another problem beyond this~-- namely the presence of an unstable ``numerical'' mode that grows without bound, leading to the lack of classical recollapse. We now derive this using von Neumann stability analysis, and show this mode does not exist for the improved quantization scheme of Ashtekar et al. For both models, we make the ansatz for the wavefunction
\[
	\Psi(m, \phi) = \psi_m e^{i \omega \phi}.
\]
Here, $m$ is a placeholder for the appropriate variable of the two models (scaled as appropriate), while $\phi$ is the value of the scalar f\/ield. For the Green--Unruh model, this is $\mu$, which is proportional to the triad component $p$; for the improved model, this is the volume $v$. With both equations in this ansatz, the quantum Hamiltonian constraint is of the form
\begin{gather*}
	C^+ (m) \psi_{m + 4} + C^0 (m) \psi_m + C^- (m) \psi_{m - 4} = \omega^2 B(m) \psi_m.
\end{gather*}
If in addition, we assume $\psi_m = \lambda^{m/4}$ in the asymptotic limit $m \to \infty$, where~$\lambda$ is a constant, then this leads to a quadratic equation in $\lambda$, namely
\[
	C^+ _0 \lambda^2 + \big(C^0 _0 - \omega^2 B_0\big) \lambda + C^- _0 = 0,
\]
where $C^+ _0$, $C^0 _0$, $B_0$ and $C^- _0$ are the limits of the appropriate function for $m \to \infty$. Using the standard formula, this gives a solution
\begin{gather*}
	\lambda = \frac{(\omega^2 B_0 - C^+ _0) \pm \sqrt{(C^0 _0 - \omega^2 B_0)^2 - 4 C^+ _0 C^- _0}}{2 C^+ _0}.
\end{gather*}
If $|\lambda| > 1$, then we have a solution corresponding to this constant which increases in magnitude without bound; this can be easily seen, since in our ansatz,
\begin{gather*}
	\lim_{m \to \infty} \left( \frac{\psi_{m+4}}{\psi_m} \right) = \lambda.
\end{gather*}
Now we look at the particular functions for each model, and f\/ind their corresponding values of~$\lambda$.

For the Green--Unruh model, we have that
\begin{gather*}
	C^+ (m)  =  [V(m + 5) - V(m + 3)] e^{-2i \Gamma \mu_0},		\\
	C^0 (m)  =  - [2 - 4 \mu_0 ^2 \gamma^2 (\Gamma^2 - \Gamma)][V(m+1) - V(m - 1)],	\\
	C^- (m)  =  [V(m - 3) - V(m  - 5)] e^{2i \Gamma \mu_0},	\\
	B(m)  =  - \frac{8 \pi G \hbar^2 (\mu_0 \gamma \ell_p)^3}{6} d_{j, l} (m)
\end{gather*}
with the volume function given by
\[
	V (m) = \left( \frac{|m| \mu_0 \gamma \ell_p ^2}{6} \right)^{3/2}
\]
the function $d_{j, l} (m)$ coming from the quantum ambiguity in the inverse volume operator~\cite{Boj04}, def\/ined as
\[
	d_{j, l} (m) = \left[ \frac{9}{j(j + 1)(2j + 1) \gamma \ell^2 _p l} \sum_{r = -j} ^j r \left( \frac{\gamma \ell^2 _p}{6}\right)^l (m \mu_0 + 2 r)^l \right]^{3/(2 - 2l)}
\]
and $\Gamma = 1/2$ indicates the closed model. For large values of $m$, then the inverse volume operator matches the classical expression closely, so that $d_{j, l} (m) \simeq (m \mu_0 \gamma \ell_P ^2 / 6)^{-3/2}$. Thus, the asymptotic expressions for the coef\/f\/icient functions are given by
\begin{gather*}
	C^+ _0 = e^{-i \mu_0} \left( \frac{\gamma \ell_P ^2 \mu_0 \sqrt{p}}{2} \right)+ O\big(m^{-1/2}\big),	\\
	C^0 _0 - \omega B_0 = - \big(2 + \gamma^2 \mu_0 ^2\big) \left( \frac{\gamma \ell_P ^2 \mu_0 \sqrt{p}}{2} \right)+ O\big(m^{-1/2}\big),	\\
	C^- _0  =  e^{i \mu_0} \left( \frac{\gamma \ell_P ^2 \mu_0 \sqrt{p}}{2} \right) + O\big(m^{-1/2}\big).
\end{gather*}
Note that the scalar f\/ield eigenvalue $\omega$ drops out, because the inverse volume dependence $B(m) \simeq m^{-3/2}$ is lower order than that of $C^0 (m) \simeq m^{1/2}$. This gives values to leading order of
\begin{gather*}
	\lambda_\pm = e^{i \mu_0} \left(1 + \frac{1}{2} \gamma^2 \mu_0 ^2 \pm \frac{\gamma \mu_0}{2} \sqrt{4 + \gamma^2 \mu_0 ^2} \right).
\end{gather*}
Thus, one eigenvalue is larger than one in magnitude; using the constant value $\mu_0 = \sqrt{3}/4$ and $\gamma = 0.2735$, then $|\lambda_+| \simeq 1.126$ while $|\lambda_-| \simeq 0.8884$.

Turning now to the improved constraint of Ashtekar et al., we have
\begin{gather*}
	C^+ (v) = \frac{3\pi KG}{8} |v+2| \big| |v+1| - |v+3| \big|,	\qquad
	C^- (v) = C^+ (v - 4),								\\
	C^0 (v) = -C^+ (v) - C^- (v) + \frac{\pi G}{2} \left\{ 3 K \left[ \sin^2 \left( \frac{{\bar \mu} \ell_0}{2} \right) - \frac{{\bar \mu}^2 \ell_0 ^2}{4} \right] |v| - \frac{\gamma \ell_0 ^2}{3} \left| \frac{v}{K} \right|^{1/3} \right\} \\
\phantom{C^0 (v) =}{}\times \big||v+1| - |v-1| \big|, \\
	B(v) = \left( \frac{3}{2} \right)^3 K |v| \big| |v+1|^{1/3} - |v-1|^{1/3} \big|^3
\end{gather*}
and ${\bar \mu} (v) = 3 \sqrt{2} / 2 (K / v)^{2/3}$. Using the same methodology as before, we have that for large magnitude of $v$,
\begin{gather*}
	C^+ _0 = C^- _0 = \frac{3 \pi K G |v|}{4} + O(1),					\qquad
	C^0 _0 - \omega^2 B_0 = - \frac{3\pi K G |v|}{2} + O\big(m^{1/3}\big).
\end{gather*}
Plugging these into the equation for $\lambda_\pm$ gives $\lambda_\pm = 1$, and the equation is stable.

\subsection*{Acknowledgements}

We would like to thank M.~Mart\'in-Benito for comments on the manuscript.
DB acknowledges f\/inancial support from Spanish Ministry of Education through
National Program No.~I-D+i2008-2011. This work has been supported by
the Spanish MICINN Project No.~FIS2008-06078-C03-03, and by Institute of
Gravitation and the Cosmos (PSU). GK acknowledges research support from NSF
Grant Nos. PHY-0831631, PHY-0902026, PHY-1016906 and PHY-1135664.

\pdfbookmark[1]{References}{ref}
\LastPageEnding

\end{document}